\documentclass[aps,prl,twocolumn,superscriptaddress,showpacs,amsmath,amssymb]{revtex4-1}
\usepackage{amsmath}
\usepackage{amsfonts}
\usepackage{amssymb}
\usepackage{graphicx}
\usepackage{color}

\usepackage{hyperref}

\begin{document}

\title{Matrix Product Renormalization Group: \\ Potential Universal Quantum Many-Body Solver}

\author{Masahiko~G.~Yamada}
\email[]{masahiko.yamada@gakushuin.ac.jp}
\affiliation{Department of Physics, Gakushuin University, Mejiro, Tokyo, 171-8588, Japan}
\affiliation{Department of Materials Engineering Science, Osaka University, Toyonaka 560-8531, Japan}
\author{Takumi~Sanno}
\affiliation{Department of Materials Engineering Science, Osaka University, Toyonaka 560-8531, Japan}
\author{Masahiro~O.~Takahashi}
\affiliation{Department of Materials Engineering Science, Osaka University, Toyonaka 560-8531, Japan}
\author{Yutaka~Akagi}
\affiliation{Department of Physics, The University of Tokyo, Hongo, Tokyo, 113-0033, Japan}
\author{Hidemaro~Suwa}
\affiliation{Department of Physics, The University of Tokyo, Hongo, Tokyo, 113-0033, Japan}
\author{Satoshi~Fujimoto}
\affiliation{Department of Materials Engineering Science, Osaka University, Toyonaka 560-8531, Japan}
\affiliation{Center for Quantum Information and Quantum Biology, Osaka University, Toyonaka 560-8531, Japan}
\author{Masafumi~Udagawa}
\affiliation{Department of Physics, Gakushuin University, Mejiro, Tokyo, 171-8588, Japan}

\date{\today}

\begin{abstract}
The density matrix renormalization group (DMRG) is a celebrated
tensor network algorithm, which computes the ground states
of one-dimensional quantum many-body systems very efficiently.
Here we propose an improved formulation of continuous tensor network algorithms,
which we name a matrix product renormalization group (MPRG).
MPRG is a universal quantum many-body solver,
which potentially works at both zero and finite temperatures,
in two and higher dimensions, and is even applicable to open
quantum systems. Furthermore,
MPRG does not rely on any variational principles and thus supports
any kind of non-Hermitian systems in any dimension.
As a demonstration, we present critical properties of
the Yang-Lee edge singularity in one dimension as a representative
non-Hermitian system.
\end{abstract}

\maketitle

\textit{Introduction}. ---
Variational principles appear everywhere in physics. For example,
classical mechanics can be formulated by the principle of
least action. Even in quantum systems, the variational principle is
the central assumption to solve various problems efficiently, which is
especially useful to obtain the ground state of Hermitian
systems.

One of the most successful applications of the quantum variational
principle is the density matrix renormalization group (DMRG) algorithm
for one-dimensional (1D) quantum many-body systems~\cite{White1992,White1993}.
DMRG is a tensor network algorithm which systematically computes the ground state
of 1D quantum many-body systems based on a variational
principle~\footnote{See e.g. Ref.~\cite{Schollwock2011} for review}.
However, due to the variational
principle, most of the methods, including DMRG, mainly deal with
Hermitian quantum systems, and it still remains
a challenging problem to develop a technique for
open quantum systems like non-Hermitian
systems~\cite{Yamamoto2022}.

Non-Hermitian systems attract attention these days due to the recent
experimental progress~\cite{Muller2012,Daley2014,Peng2015,Sieberer2016,Ashida2020}.
While the development in the realization of open
quantum systems or engineering non-Hermitian terms in cold atomic systems is remarkable,
the dissipation effect is ubiquitous and inevitable in any kind of realistic
quantum systems, and thus dealing with non-Hermitian dissipation terms should be
at the heart of the application of quantum mechanics to the real world.

Variational principles are not directly applicable to non-Hermitian quantum
systems, and thus solving interacting open quantum many-body systems
has long been a very difficult task. While non-variational method like
infinite time-evolving block decimation (iTEBD)~\cite{Vidal2007,Orus2008}
has been used to simulate non-Hermitian 1D quantum many-body systems~\cite{Ashida2017},
it requires a large bond dimension to obtain a reliable result from iTEBD.
We need a more efficient algorithm which can solve non-Hermitian systems with
a direct method.

However, obtaining the direct solution of the many-body Schr\"odinger equation
without a variational principle is highly challenging. One of the main reasons
is the numerical instability of solving the generalized eigenvalue problem.
We find that this instability can be removed by using the so-called \textit{continuous}
tensor product state~\cite{Verstraete2010,Hastings2015,Tirrito2018,Tang2020},
which can be regarded as a hyper-dual-number-coefficient tensor network
state~\footnote{In this Letter, we restrict the definition of continuous tensor
networks to those where only the imaginary time direction is made continuous.
Spatial directions are always discrete and should be treated differently}.
This formalism allows us to diagonalize general many-body Hamiltonians efficiently
based on tensor network ansatz for both Hermitian and non-Hermitian quantum systems
in a nonvariational way.
Indeed, the use of hyper-dual numbers drastically simplifies the algorithm,
and facilitates the implementation with modern programming languages.
We refer to a continuous tensor network algorithm with this hyper-dual number
eigensolver as a matrix product renormalization group (MPRG).

Compared with other methods, including a quantum Monte Carlo, MPRG has
the following three advantages:
(i) the absence of numerical sign problems, (ii) the absence of Trotter errors,
and (iii) the absence of finite-size effects. MPRG is a potential universal
quantum many-body solver, which can compute the correct thermodynamic quantities
in the large-bond-dimension limit.

In this Letter, we propose MPRG as a universal quantum many-body solver.
MPRG potentially works at both zero and finite temperature, in any dimensions,
and is even applicable to open quantum systems without a variational principle.
As a demonstration, we formulate MPRG in the 1D case, and show that MPRG can solve
various types of many-body problems very efficiently in both Hermitian and non-Hermitian
cases. Specifically, we present a benchmark result for the Yang-Lee edge singularity
of the complex-valued-transverse-field Ising model~\cite{YangPR87,LeeOR87,FisherPRL40}.
Furthermore, the extension to two and higher dimensions is straightforward.
This will eventually give rise to an ultimate solution to notorious numerical
sign problems.

\textit{Continuous matrix product state and continuous projected entangled-pair states}. ---
For 1D chain systems, the tensor product state ansatz is usually called a matrix product state
(MPS). MPS is used in many quantum many-body solvers, such as DMRG and iTEBD.
Those algorithms use a discrete version of MPS, which is very efficient due to the
existence of the so-called canonical form. The generalization to higher dimensions
is called projected entangled-pair states (PEPS), and again it is useful to solve
the ground states of various quantum many-body systems at zero temperature.

Recently, a continuous variant of tensor product states appeared, and it is found that
the continuous version is useful, especially for finite-temperature systems.
For example, by using a continuous MPS (cMPS), we can eliminate the error
associated with the Trotter decomposition, leading to
an accurate computation of the partition function at finite temperature. Later
we will see that cMPS can be reformulated as a hyper-dual-number-coefficient MPS
in cases we are interested in, and
this view will simplify various technical difficulties in the calculation.
Similarly, we define a continuous PEPS (cPEPS) as a hyper-dual-number-coefficient
PEPS for higher dimensions~\footnote{The use of hyper-dual numbers is only sufficient
for systems with only one- or two-body interactions. If there is a three-body interaction,
a higher object, like hyper-hyper-dual numbers, will be required.}.

\textit{Matrix product renormalization group}. ---
A hyper-dual number is an extended version of a dual number. We have two elements
$\varepsilon_1$ and $\varepsilon_2$ that satisfy $\varepsilon_1^2 = 0$, $\varepsilon_2^2 = 0$,
and $\varepsilon_1$ commutes with $\varepsilon_2$. These relations are enough to regard
$\varepsilon_1$ and $\varepsilon_2$ as two independent infinitesimal values.
Any hyper-dual number can be denoted by $a+b\varepsilon_1 + c\varepsilon_2 + d\varepsilon_1\varepsilon_2$,
where $a$, $b$, $c$, and $d$ are complex numbers, and using this quantity,
we can compute the derivative of functions up to the second order.

Similarly, the Taylor expansion about $\sqrt{\Delta\tau}$ can be possible up to
the second order, where $\Delta\tau$ is a Trotter error. From a simple fact
$[(\varepsilon_1 + \varepsilon_2) / \sqrt{2}]^2=\varepsilon_1\varepsilon_2$,
we can easily identify $(\varepsilon_1 + \varepsilon_2) / \sqrt{2}=\sqrt{\Delta\tau}$
and $\varepsilon_1\varepsilon_2=\Delta\tau$ to translate the continuous tensor network
into the hyper-dual-number-coefficient tensor network.

In the appendix of Ref.~\cite{Tang2020}, many continuous matrix product operators (cMPOs)
for various quantum models are provided, and we can easily translate
them into hyper-dual-number-coefficient MPOs. For example, cMPO for
the transverse-field Ising model is
\begin{align}
    M &= \begin{pmatrix}
        I + \Gamma S^x & \sqrt{J / 2}(\varepsilon_1 + \varepsilon_2) S^z \\
        \sqrt{J / 2}(\varepsilon_1 + \varepsilon_2) S^z & 0
    \end{pmatrix},
\end{align}
where $M$ is a rank-4 cMPO, $S^{x,y,z}$ is the spin-$1/2$ operator, and
$I$ and $0$ are identity and zero matrices, respectively. The parameters
$J$ and $\Gamma$ are model parameters for the transverse-field Ising model
defined later. After this identification, we still need to make a contraction
of these infinite tensor networks. However, we will see that the contraction
is much easier than conventional discrete tensor networks, thanks to the
introduction of hyper-dual numbers.

To do this contraction with hyper-dual numbers, we propose another type of
continuous tensor network method named MPRG.
Let us discuss the ground state calculation of MPRG. The algorithm
for the ground state is similar to infinite DMRG (iDMRG), or the variational
uniform matrix product state (VUMPS) ansatz~\cite{ZaunerStauber2018}. The difference
lies in the fact that the transformation to a canonical form of MPS is no longer necessary,
and cMPS environmental tensors can be computed directly in a translation-invariant
form. While we need to solve the generalized eigenvalue problem in this case,
for hyper-dual-number-coefficient cMPS, it is enough to compute the
\textit{derivative} of the generalized eigenvalue problem. The
derivative of the generalized eigenvalue problem can be computed
directly with a forward-mode automatic differentiation, or we can
simply use the power method. We found that the power method has a better
performance in both Hermitian and non-Hermitian cases to \textit{avoid}
solving the full generalized eigenvalue problem. This is the basic
idea of MPRG.

The generalization to the finite temperature is straightforward. We can
even use the \textit{same} environmental tensor calculated for the ground
state. This is a remarkable fact because it means that a single simulation
for the $T \to 0$ limit is enough for all physical temperature $T$. This is
due to the generalized Hellmann-Feynman theorem for continuous tensor
networks~\footnote{This form of the Hellmann-Feynman theorem generalized
to higher-order derivatives has been proven in Ref.~\cite{Tang2020} up to the second order.
The extension of this proof to higher orders is straightforward and will
be discussed in a future publication.}. Indeed, from this theorem, it
is easy to show that the free energy does not depend on $\beta = 1/T$
\textit{implicitly}, and it means that the $\beta$ dependence of the
environmental tensor does not affect any thermodynamic quantities computed
from the free energy.

We note that all the calculations in this work have been done using
the Julia language~\cite{Julia2017} with a forward-mode automatic differentiation
library Handagote.jl~\footnote{\url{https://github.com/MGYamada/Handagote.jl}}.

\textit{Fixed point iteration}. ---
Based on this observation, we can easily formulate the fixed point iteration
for the cMPS environmental tensor. The fixed point iteration for MPRG is
similar to that for VUMPS, which is the sequence of the iteration of the environment calculation,
the power method for hyper-dual-number-valued matrices, and the cMPS tensor update.

First, the environment calculation is done simply by diagonalizing
the $\varepsilon_1\varepsilon_2$-part of the hyper-dual-number-coefficient
transfer matrix. This is justified by the fact that after the convergence,
the transfer matrix always becomes $I - \varepsilon_1\varepsilon_2 H_\mathrm{eff}$,
where $I$ is an identity matrix and $H_\mathrm{eff}$ is an effective Hamiltonian.
Thus, it is enough to diagonalize the $\varepsilon_1\varepsilon_2$-part of the transfer matrix,
and environments are always pure complex tensors; In the degenerate perturbation theory,
the first-order perturbation affects only the zeroth-order part of the wavefunction.
This fact is not altered for non-Hermitian systems, i.e., the same method can be
applied to non-Hermitian systems.

Next, the power method is used to solve the generalized eigenvalue problem
in terms of hyper-dual numbers. In order to solve the problem of the form
$A\psi = \lambda B\psi$, where $(A, B)$ is a given matrix pencil and
$(\lambda, \psi)$ is a resulting eigenpair, we compute $B^{-1}$ first
using the derivative of the matrix inverse. The eigenvector $\psi$ with the largest eigenvalue
is computed by applying $B^{-1} A$ many times, or more simply by
solving some consistency equation derived from the fixed point iteration.
We found that the latter is better for our simulation. The details of this
approach will be discussed in a future publication.

Finally, When updating the cMPS environmental tensor, it is important to stabilize the
whole fixed point iteration by the following important modification. If the
fixed point iteration from the old tensor $E_i$ to the new tensor $E_{i+1}$
has the form of $E_{i+1} = f(E_i)$, then the modified (stabilized) version
of this fixed point iteration should be $E_{i+1} = (1-\alpha) E_i + \alpha f(E_i)$.
Here $\alpha \in (0, 1]$ is a small number chosen carefully to stabilize the
fixed point iteration. The optimal $\alpha$ strongly depends on the model
and the parameter, and we can only know its value empirically.

\textit{Comparison with an exact solution in Hermitian systems}. ---
After the above modification, we succeed in stabilizing the calculation for
various 1D Hermitian systems. As the simplest benchmark, we demonstrate
a simulation for the 1D transverse-field Ising model, and compare the result
with the exact one. The Hamiltonian is
\begin{align}
    H_\mathrm{Ising} = -\sum^N_{j=1} \left( J S^z_j S^z_{j+1} + \Gamma S^x_j \right),
\end{align}
where $S^{x,y,z}_j$ is the spin-$1/2$ operator at site $j$ and $N$ is the system size.
Here, we impose periodic boundary conditions, and set $\lambda \equiv J/2 \Gamma$ with $\Gamma=1$.
The parameter $\lambda$ is normalized so that $\lambda=1$ corresponds to the critical point,
and we use this normalization from now on.

While the calculation for gapped cases $\lambda \neq 1$
converges very fast and accurately as by DMRG and VUMPS, the most remarkable thing of MPRG
is the accuracy at the critical point $\lambda=1$. First, we show the ground state
energy calculated for various bond dimensions $\chi$ by MPRG at $\lambda=1$, and
it converges very rapidly to the exact value calculated for the $N=10^7$ system,
as shown in Table~\ref{tab:E}.
\begin{table}[htbp]
    \centering
    {\tabcolsep = 0.5cm
    \begin{tabular}{|c|c|} \hline
    $\chi$ & $E_G/N$ \\ \hline
    $2$ & $-0.63627091$ \\ \hline
    $4$ & $-0.63659594$ \\ \hline
    $8$ & $-0.63660486$ \\ \hline
    exact ($N=10^7$) & $-0.63661982$ \\ \hline
    \end{tabular}
    }
    \caption{Ground state energy $E_G/N$ for the transverse-field Ising model at $\lambda=1$.}
    \label{tab:E}
\end{table}
The calculation with $\chi=8$ is already very accurate, and this means that MPRG has
the ability to represent a gapless state with a very small bond dimension.

\begin{figure}
    \includegraphics[width=8cm]{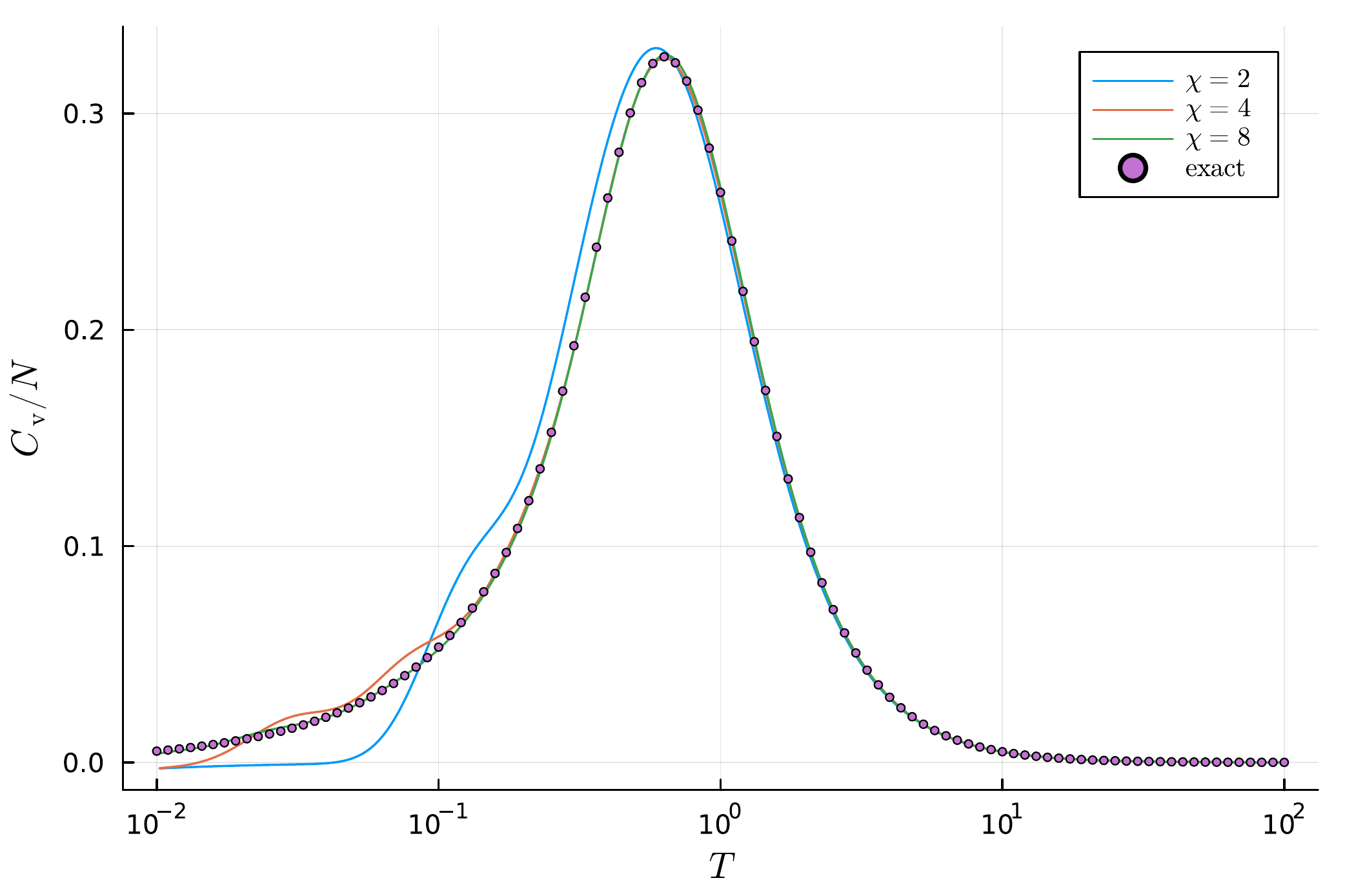}
    \caption{Specific heat for the 1D transverse-field Ising model calculated by MPRG (solid lines)
    for different bond dimensions $\chi$. The exact values ($N=10^4$) are shown in purple dots.}
    \label{fig:ising}
\end{figure}

The high representation ability of MPRG is more evident in the finite-temperature
calculation. As shown in Fig.~\ref{fig:ising}, the heat capacity calculated for $\chi=8$
by MPRG, shown in the green line, is almost converged to the exact result ($N=10^4$) shown
in purple dots. While we see an oscillation at low temperature for small bond dimensions
$\chi=2,\, 4$, the accuracy of $\chi=8$ is remarkable as we see the correct gapless behavior
at the lowest temperature. Although MPS methods usually fail to capture the gapless
behavior~\cite{Huang2015}, it is shown that cMPS is more suitable for gapless (critical) systems.

\textit{Demonstration in Non-Hermitian systems}. ---
Next, in order to demonstrate the potential of MPRG, we simulate a more exotic critical
behavior, which has never been discussed in previous variational approaches. As a benchmark
result for non-Hermitian systems, we reproduce the phase diagram of the Yang-Lee Ising spin model.
The Yang-Lee Ising spin model is the 1D transverse-field Ising spin model with an additional
pure imaginary magnetic field $h$:
\begin{align}
    H_\mathrm{YL} = -\sum^N_{j=1} \left( J S^z_j S^z_{j+1} + \Gamma S^x_j + ih S^z_j \right).
\end{align}
We also impose periodic boundary conditions on the system.
The magnetization of this model exhibits singular behavior with a negative exponent close to
the critical magnetic field $h_c$. This \textit{non-unitary} critical behavior is called Yang-Lee edge singularity~\cite{YangPR87,LeeOR87,FisherPRL40}.
This critical phenomenon can be described by the \textit{non-unitary} conformal field theory (CFT) with a central charge $c=-22/5$~\cite{CardyPRL54}.
This model also possesses the \textit{Parity-Time} $(\mathcal{PT})$ symmetry, which imposes a constraint on the spectrum of this model.
From the eigenvalue spectrum theory, the spectrum is real when the ground state is invariant with the $\mathcal{PT}$ symmetry,
while after breaking the $\mathcal{PT}$ symmetry, the spectrum changes to complex conjugate pairs~\cite{BenderPRL80, BenderPRD87}.
Therefore, the Yang-Lee model attracts much attention from viewpoints of the \textit{non-unitary} CFT and as the $\mathcal{PT}$-symmetric system.
Notably, elementary excitations close to $h_c$ could be \textit{non-unitary non-Abelian anyons} and satisfy the same fusion rule
as that of Fibonacci anyons~\cite{ArdonneIOP13,FreedmanPRB85,Sanno2022}.

\begin{figure}
    \includegraphics[width=8cm]{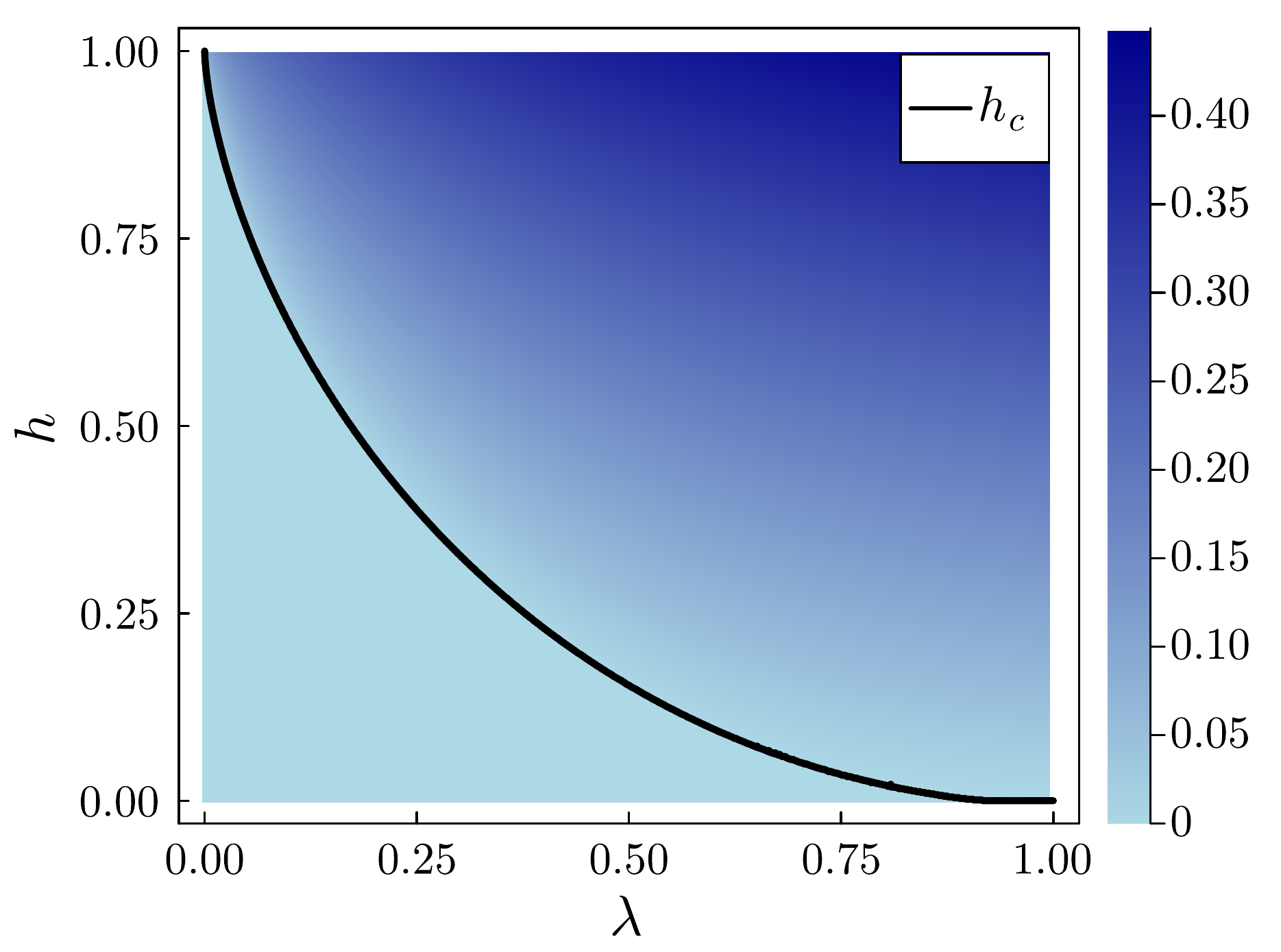}
    \caption{Imaginary part of $E_G/N$. These values are numerically zero up to $h_c$.}
    \label{fig:phase_yang-lee}
\end{figure}

Breaking the $\mathcal{PT}$ symmetry, the spectra of the Yang-Lee model drastically change from real to complex spectra.
The imaginary part of the spectra directly corresponds to the phase diagram of the Yang-Lee model.
Fig.~\ref{fig:phase_yang-lee} shows the imaginary part of the ground state energy par site $E_G/N$~\footnote{Here
the ground state means that the real part of the energy is at its minimum.}.
The line in Fig.~\ref{fig:phase_yang-lee} means the critical magnetic field $h_c$.
\begin{table}[htbp]
    \centering
    {\tabcolsep = 0.5cm
    \begin{tabular}{|c|c|} \hline
    $\lambda$ & $h_c$ \\ \hline
    $0.10$ & $0.636$ \\ \hline
    $0.20$ & $0.457$ \\ \hline
    $0.30$ & $0.328$ \\ \hline
    $0.40$ & $0.230$ \\ \hline
    \end{tabular}
    \begin{tabular}{|c|c|} \hline
        $\lambda$ & $h_c$ \\ \hline
        $0.50$ & $0.154$ \\ \hline
        $0.60$ & $0.095$ \\ \hline
        $0.70$ & $0.052$ \\ \hline
        $0.80$ & $0.021$ \\ \hline
    \end{tabular}
    }
    \caption{Critical magnetic field $h_c$.}
    \label{tab:h_c}
\end{table}
Table~\ref{tab:h_c} shows some results of the critical magnetic field $h_c$.
This numerical result is consistent with the previous study based on the exact diagonalization method~\cite{GehlenIOP24}.

\begin{figure}
    \includegraphics[width=8cm]{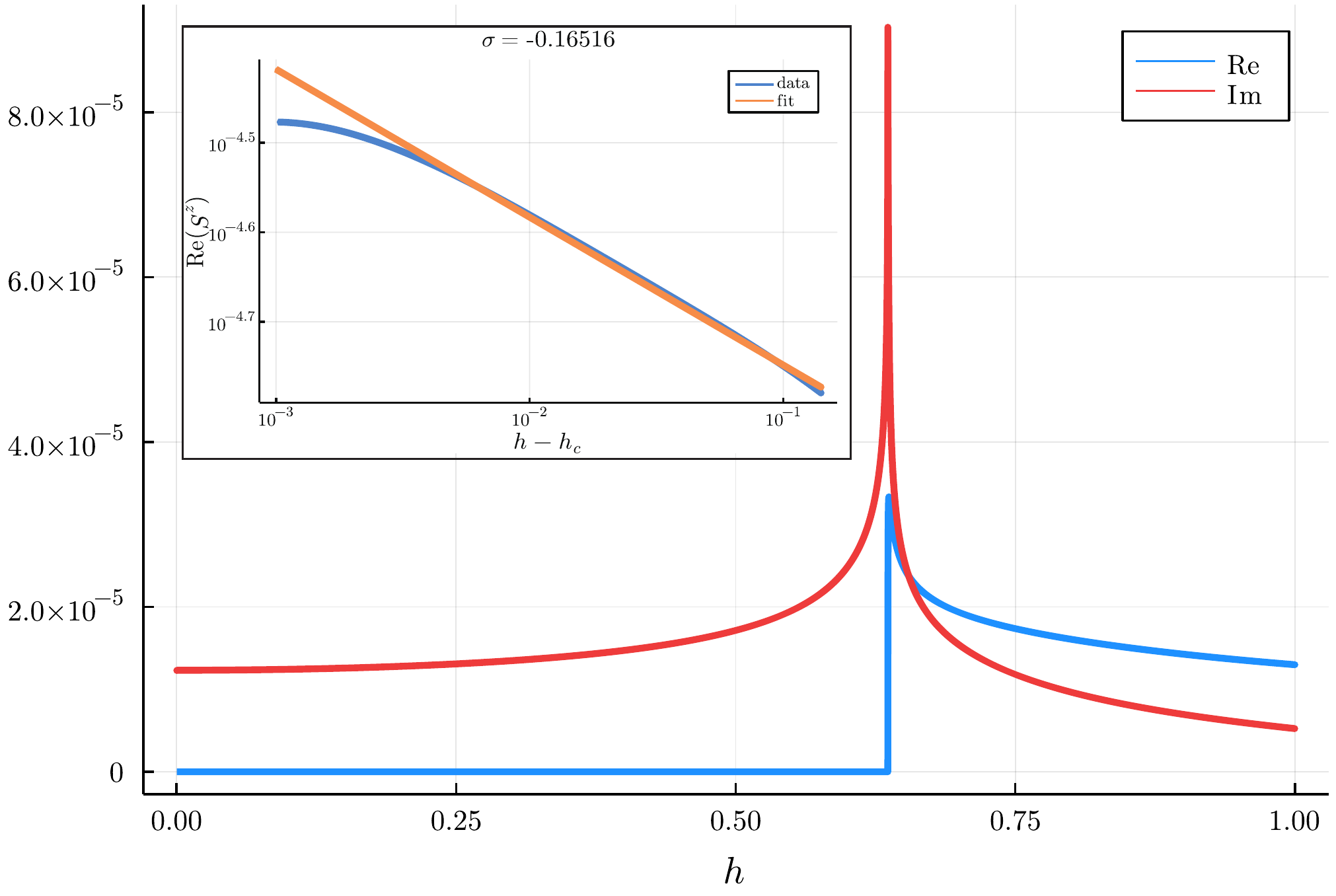}
    \caption{The magnetization $S^z$ as a function of $h$ at $\lambda=0.1$. 
    The inset shows the log-log plot of the real part of $S^z$ for $h> h_c$ region with a power function fitting.}
    \label{fig:mag_yang-lee}
\end{figure}

Moreover, we would like to discuss the magnetization of the Yang-Lee model as a characteristic feature of the \textit{non-unitary} critical phenomena.
From the renormalization group method and the \textit{non-unitary} CFT, the exponent $\sigma$ of the magnetization, i.e.,
$\langle S^z \rangle\sim (h-h_c)^{\sigma}$, is $-1/6$ in the thermodynamic limit.
Fig.~\ref{fig:mag_yang-lee} shows the magnetization of the Yang-Lee model.
The real part of the magnetization drastically changes from zero to a finite value by breaking the $\mathcal{PT}$ symmetry.
These numerical results would imply that the exponent of the magnetization is equal to a negative value as a function of the magnetic field $h$.
In fact, by breaking the $\mathcal{PT}$ symmetry, we succeed in estimating the component as $\sigma=-0.165(2)$ as shown in the inset of Fig.~\ref{fig:mag_yang-lee}
in the case $\lambda=0.1$, which is consistent with the exact value.
Note that, through our calculations for non-Hermitian systems, the bond dimension is set $\chi=2$, which indicates that
the computational cost is low enough.

\textit{Conclusion}. ---
We have proposed a universal quantum many-body solver MPRG, and MPRG supports
infinite systems, finite temperature, and non-Hermitian systems.
We have studied the Yang-Lee edge singularity of the
complex-valued-transverse-field Ising model to show the high
potential of this algorithm. The representation ability is remarkable,
as it can accurately represent the criticality described
by the \textit{non-unitary} CFT.

Finally, we should mention that the extension of MPRG to two or
higher dimensions is straightforward by using cPEPS.
We will present more systematic studies for various quantum systems
in a future publication.

\begin{acknowledgments}
We thank T.~Okubo, F.~Pollmann, W.~Tang, H.-H.~Tu, and L.~Wang for
fruitful discussions. M.G.Y. is supported by JST PRESTO Grant No. JPMJPR225B.
T.S. and M.O.T. are supported by a JSPS Fellowship for Young Scientists.
Y.A. is supported by JST PRESTO Grant No. JPMJPR2251.
S.F. is supported by JST CREST Grant No. JPMJCR19T5, Japan.
This work was supported by JSPS KAKENHI Grant Nos. JP20H05655, JP20K14411, JP20J20468,
JP21H01039, JP22K03508, JP22K14005, JP22H01147, and JP22J20066, and JSPS Grant-in-Aid
for Scientific Research on Innovative Areas ``Quantum Liquid Crystals'' (KAKENHI Grant No.
JP22H04469).
\end{acknowledgments}

\bibliography{paper}

\end{document}